\documentclass[traditabstract]{aa}

\usepackage{graphicx}
\usepackage{txfonts}
\usepackage{lipsum}
\usepackage{subcaption}         
\usepackage{lscape}             
\usepackage{placeins}           
\usepackage{makecell}
\usepackage{xcolor}
\definecolor{refblue}{RGB}{50,50,170}  

\usepackage[colorlinks=true,
    citecolor=refblue,
    linkcolor=refblue,
    urlcolor=refblue]{hyperref}
\usepackage{orcidlink}

\begin{document}

\title{The effect of  triples on the Chemically Homogeneous Evolution Channel for producing Gravitational Wave Sources}
\titlerunning{The effect of  triples on the Chemically Homogeneous Evolution Channel}

\author{Silvia A. Popa\orcidlink{0009-0006-3186-5826}\inst{1}\corrauth{spopa@mpa-garching.mpg.de}
    \and Selma E. de Mink\orcidlink{0000-0001-9336-2825}\inst{1}
    \and Stephen Justham\orcidlink{0000-0001-7969-1569}\inst{1}
    \and \\ Jakob Stegmann\orcidlink{0000-0003-2340-8140}\inst{1}
    \and Alejandro Vigna-Gomez\orcidlink{0000-0003-1817-3586}\inst{2,1}
    \and Lieke A. C. van Son\orcidlink{0000-0001-5484-4987}\inst{3}}

\institute{Max Planck Institute for Astrophysics, Karl-Schwarzschild-Str.~1, 85748 Garching, Germany
\and Niels Bohr International Academy, Niels Bohr Institute, Blegdamsvej 17, DK-2100 Copenhagen, Denmark
\and Department of Astrophysics, IMAPP, Radboud University Nijmegen, PO Box 9010, 6500 GL Nijmegen, The Netherlands}

\date{Received ... }

\abstract
{Despite the growing number of gravitational-wave (GW) detections, the formation of merging binary black holes (BBHs) is still debated. One of the proposed formation channels is Chemically Homogeneous Evolution (CHE), which makes distinct predictions for the properties of the resulting black holes (BHs): high masses ($\gtrsim35\, M_\odot$) and large, orbitally-aligned spins. These predictions are based on studies that consider this channel only in isolated binaries. However, most massive stars reside in triples. We investigate how a tertiary affects the CHE channel. We use detailed binary models computed with MESA for the inner CHE systems coupled with analytical prescriptions for the hierarchical triple dynamics. Specifically, we study the von Zeipel-Lidov-Kozai mechanism that can accelerate the merger of the inner BBHs through high-eccentricity excitations. 

We show that a tertiary companion enables mergers at higher metallicities, where the classical binary CHE channel does not operate. At $Z_\odot/3$, triples produce BHs with lower masses (around $15\,M_\odot$) and lower spins (around 0.25) compared to the low-metallicity classical scenario. We further expect triple dynamics to lead to spin-orbit misalignment and potential residual eccentricities at merger. In triples at $Z_\odot/3$, half of the mergers occur within 1 Gyr. Moreover, 10\% of the population leads to prompt mergers within 1 Myr (1\% within 1 kyr) that are of potential interest for electromagnetic counterparts. We conclude that the triple CHE channel is promising as (i) it may contribute more substantially to the local merger rate than had been expected for the classical CHE channel due to the shorter delay times and effectiveness at higher metallicities and (ii) produces mergers with properties that more closely resemble the lower masses and lower spins inferred for most of the observed GW merger population. 
}


\maketitle

\nolinenumbers

\section{Introduction}

The rapidly growing number of gravitational-wave (GW) detections, which has reached several hundred at the time of writing, provides a powerful dataset for investigating the population of merging binary black holes \citep[BBHs,][]{gwtc5}. Several formation channels have been proposed to explain the observed mergers, ranging from isolated binary evolution to triple systems and dynamical interactions in dense environments, but their origin is still debated \citep{Abbott_2016, Mapelli_2021, Mandel_2022}. 

Observations indicate that a large fraction of massive stars, which are the progenitors of the GW events, reside in triples \citep{Sana_2014, Bordier_2022, Offner_2023, Bordier_2024, Frost_2025}, with fractions as high as 70\% for massive ($> 16\,M_\odot$) O-type stars \citep{Moe_2017}.  
In hierarchical triples, the tertiary induces oscillations in the eccentricity and inclination of the inner binary system through the von Zeipel-Lidov-Kozai (ZLK) mechanism \citep{Zeipel_1910, Lidov_1962, Kozai_1962}. The periodic excitation of high eccentricities enhances gravitational-wave emission, shrinking the binary orbit and reducing the coalescence time of the inner BHs. Various studies have therefore explored the consequences of triples in the context of GW studies \citep{Silsbee_2017,Antonini_2017,Antonini_2018,Grishin_2018,Liu_2018,Rodriguez_2018,Mangipudi_2022,Stegmann_2022, Dorozsmai_2023,Dorozsmai_2026, Vigna_2025,  Stegmann_2026, stegmann2026massspinpropertiesblackhole}.

For very tight binaries, which are the focus of this study, the triple fraction might be even higher \citep{Moe_2017, Fabrycky_2007, Bataille_2018, Kummer_2023, Moe_2026}. For solar-type spectroscopic binaries, \cite{Tokovinin_2006} showed that 96\% (consistent with 100\%) of the systems with periods $P<3\, \text{days}$ have a triple companion. Similarly high fractions were also found by \cite{Rucinski_2006} for W UMa systems, a type of low-mass main-sequence contact binaries. Tertiary companions have also been found for high-mass contact binaries \citep[e.g.,][]{Kennedy_2010, Mayer_2013, Lorenzo_2017, Janssens_2021}. 

The channel we consider in this work is the Chemically Homogeneous Evolution (CHE) channel \citep{Mandel_2016, Marchant_2016, Mink_2016}. It involves two initially massive ($\gtrsim40\,M_\odot$) stars in a very tight binary orbit ($P \lesssim 3$ days), where tidal locking forces both stars to rotate rapidly \citep{DeMink_2008, DeMink_2009, Song_2016}, see also \citet{Paczynski_1976}. First discussed in the context of single-star evolution, rapidly rotating stars develop steady-state currents, known as Eddington-Sweet circulation, which efficiently mix the stellar interior by transporting helium to the surface and fresh hydrogen to the core \citep{Vogt_1925, Eddington_1925, Sweet_1950}. Mixing can be further enhanced in close binaries through tidal deformation \citep{Hastings_2020} and shear resulting from tidal synchronization \citep{Song_2013}, preventing the build-up of a chemical gradient and leading to CHE. Chemically homogeneous evolving stars do not experience the dramatic expansion in stellar radius after the main sequence that is expected from normal stellar evolution \citep{Maeder_1987, Yoon_2006}. Instead, they stay compact while they evolve into helium stars and eventually collapse into black holes. In a binary system, CHE stars may get in contact but avoid major mass transfer and common envelope episodes and naturally form black holes that merge within the age of the Universe owing to their initial high masses and tight orbits \citep{Marchant_2016, Mandel_2016, Mink_2016}. 

The CHE channel, in its classical formulation, makes distinct predictions for the properties of the resulting black holes (BHs): nearly equal component masses of about $30\,M_\odot$ and above, and high spins aligned with the orbital angular momentum \citep{Marchant_2016, Mandel_2016, Mink_2016}. It has therefore been speculated that the CHE channel is a potential contributor to the feature observed around $35\, M_\odot$ in the BBH mass distribution from GW detections \citep[][and references therein]{Li_2025, Kishore_2025}. However, the expectation that the CHE channel produces high aligned BH spins is in tension with the overall GW population. While there are a few individual events with rapidly rotating BHs that are potential candidates for the CHE channel \citep{Qin_2022, Stegmann_2025a, Popa_2025}, the observed population is dominated by low-spin systems, with \makebox{$\sim85$\% of BHs} from the latest data release showing individual dimensionless spin magnitudes below 0.25 \citep{gwtc5}.

The ubiquity of triples especially in short-period systems, has motivated studies that investigate the impact of triple dynamics with a CHE inner binary. In a semi-analytical study, \cite{Vigna-Gomez_2021} discussed tight triple systems with a CHE inner binary as potential progenitors of unusually massive GW events. However, their analysis was restricted to coplanar systems, neglecting three-body effects such as the ZLK mechanism. 
The role of three-body dynamics was further explored by \cite{Dorozsmai_2023} by incorporating a simplified recipe to mimic CHE binaries into the rapid population synthesis code \texttt{TRES} \citep{Toonen_2016} following the approach by \cite{Riley_2021}. They explore different evolutionary scenarios and show that the tertiary companion can dynamically assist a merger in the inner CHE binary. More recently, \cite{Vigna_2025} studied the ZLK mechanism in triples using a detailed stellar evolution model for one representative inner CHE binary system at two different metallicities ($Z= 0.00042 \text{ and } 0.0042$). Including short-range forces that could quench the eccentricity excitations, they explore different triple configurations, showing that some of them lead to very prompt mergers of the inner BBHs.

In this work, we use a grid of detailed MESA stellar evolution models of the inner CHE binary, coupled with analytical prescriptions for the hierarchical triple dynamics following \citet{Vigna_2025} in order to investigate the properties of the BBHs formed through the CHE channel in triples. While the isolated binary CHE channel is limited to low metallicities, we focus on a regime at $Z_\odot/3$, where the CHE channel is not efficient in producing BH mergers from isolated binaries \citep{Buisson_2020}. At these higher metallicities, mass loss due to stellar winds widens the orbit to the extent that the resulting BHs are too far apart to merge within a Hubble time. However, a tertiary can bring the BHs closer and reduce the merger time, opening up the CHE channel to higher metallicities and potentially changing the predictions of the isolated CHE channel.

This paper is organized as follows. In Section~\ref{sec:methods} we describe the grid of detailed stellar models and the treatment of triple dynamics. In Section~\ref{sec:results} we explore the parameter space across different inner CHE binary systems and triple configurations that lead to mergers that would have not been possible in isolated binaries. In Section ~\ref{sec:4} we compare predictions for the BH merger populations from the CHE triple and binary channels, at two fixed metallicities. We discuss caveats of our work in Section~\ref{sec: discussion} and summarize our results in Section~\ref{sec:conclusion}.

\section{Methods}
\label{sec:methods}

\subsection{CHE models}
\label{sec:che_models}

\subsubsection{Physics and assumptions implemented in MESA}

We use the grid of chemically homogeneously evolving (CHE) binaries from \cite{Buisson_2020}, which consists of detailed binary evolution models, extending the grid presented and discussed by \citet{Marchant_2016}. These models are computed with the one-dimensional stellar evolution code MESA, version 11701 \citep{Paxton_2011, Paxton_2013, Paxton_2015, Paxton_2018, Paxton_2019, Jermyn_2021}. All models assume that the progenitor stars have equal masses. This choice is motivated by the findings of \cite{Marchant_2016} that most CHE systems initially undergo a stable overcontact phase, during which repeated mass transfer equalizes the component masses. We briefly summarize the physics and main assumptions implemented in MESA. For a more detailed description, we refer the reader to \cite{Marchant_2016} and \cite{Buisson_2020}. 

Convection is modeled using the standard mixing-length theory \citep{Bohm-Vitense_1958}, adopting the Ledoux criterion with a mixing-length parameter \makebox{$\alpha =1.5$} and semiconvection with an efficiency parameter $\alpha_\text{sc} = 1.0$ \citep{Langer_1983}. Convective core overshooting is included during core hydrogen burning with a step function of $\alpha_{\text{OV}} = 0.345$ following \cite{Brott_2011}, which is also consistent with large modern surveys of O-type stars \citep{Lechien_2026}. Rotationally induced mixing is included through Eddington-Sweet circulation, secular and dynamical shear instabilities, and the Goldreich-Schubert-Fricke instability, with an overall efficiency parameter of $f_c = 1/30$ \citep{Pinsonneault_1989, Chaboyer_1992, Heger_Langer_Woosley_2000, Brott_2011}. Angular momentum transport due to magnetic fields is modeled using the Spruit-Tayler dynamo \citep{Spruit_2002}. Stellar winds are implemented following \citet{Yoon_2006} with mass-loss rates as in \citet{Vink_2001} for hydrogen-rich stars and \citet{Hamann_1995} (reduced by a factor of 10) for hydrogen-poor stars with a metallicity scaling of $(Z/Z_\odot)^{0.85}$ \citep{Mokiem_2007}. Although updated wind prescriptions have been developed \citep[e.g.][]{Sabhahit_2023, Bjorklund_2021, Bjorklund_2023, Higgins_2023}, they were not available at time the grids were computed. For rapidly-rotating stars, rotationally enhanced mass loss is included to ensure that the rotation rate remains below the critical value set by the balance of gravity, centrifugal forces and radiation pressure \citep{Heger_2000}. 

Binary interactions include tidal interactions following the formalism of \cite{Zahn_1975, Zahn_1977} for stars with radiative envelopes as in \cite{Hurley_2002}. Some very tight binaries might go through an overcontact phase where both stars are overflowing their Roche volumes. For the equal-mass inner binaries considered here, the systems evolve on a nuclear timescale unless they overflow the outer Lagrangian point L2, in which case strong angular momentum loss is commonly thought to trigger a common-envelope-like inspiral leading to a merger \citep[see, e.g.][]{Paczynski_1976, Flannery+1977, Pejcha_2017, MacLeod+2018, Henneco_2023}. 

The microphysics includes CO-enhanced opacities from the OPAL project \citep{Iglesias_1996}, computed using solar-scaled abundances based on \cite{Grevesse_1996}. For the reaction rates, different reaction networks were used for different evolutionary stages: an 8-isotope network for H and He burning (\texttt{basic.net}), a 9-isotope network  for C and O burning (\texttt{co\_burn.net}), and a 21-isotope network for the later evolutionary phases (\texttt{approx21.net}).

\subsubsection{Properties of the CHE grid}

The CHE grid spans the parameter space in initial primary mass ($M_{1,2}=25$ to $500  \, \mathrm{M}_\odot$), orbital period \makebox{($P_i=0.4$ to $4.0$\,days)}, and metallicity ($\log_{10}(Z) = -5.0$ to $-2.3785$), where $Z$ represents the mass fraction of elements heavier than helium. Throughout this work, we express metallicities relative to a solar value of \makebox{$Z_\odot = 0.0142$}, following \citet{Asplund_2021}; we note that the models were computed using an older value from \makebox{\citet{Grevesse_1996}}. 

\begin{figure}
    \centering
    \includegraphics[]{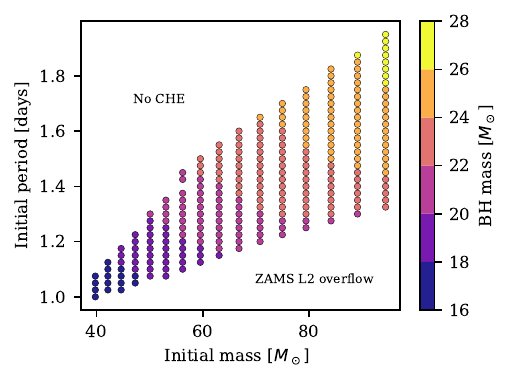}
    \caption{Overview of the chemically homogeneously evolving (CHE) models at a metallicity of $\log{Z} = -2.375 \approx Z_\odot/3$, shown as a function of their initial period and mass, colorcoded by the final black hole (BH) mass. Binary models that are in a sufficiently tight configuration experience overflow of the outer Lagrangian point L2 on the zero-age main sequence (ZAMS) and are expected to merge. For too wide systems, tidal synchronization becomes inefficient and the models do not undergo CHE. For clarity, we only show systems with masses below $100\, M_\odot$.}
    \label{fig:period_mass}
\end{figure}

Figure \ref{fig:period_mass} shows the CHE models at the highest metallicity in the grid, $Z_\odot/3$. Binary systems that start in a sufficiently tight orbital configuration on the zero-age main sequence (ZAMS) overflow their outer Lagrangian point L2 and they are expected to merge. On the other hand, if the system is too wide, tidal synchronization becomes inefficient and the stars do not rotate fast enough for CHE to take place. These systems are not evolved further. Only CHE models, where the chemical composition between the core and the surface does not differ by more than 20\%, are evolved until core carbon depletion. The final fate of each star is inferred based on its core properties, using the assumption that stars in the considered mass range undergo direct collapse to form BHs. Under the assumption of total mass and angular momentum conservation, we use the mass and the progenitor's spin at core carbon depletion as proxies for the mass and spin of the resulting BHs. At this metallicity, the stars lose more than half of their mass through stellar winds, leading to slower-spinning, lower-mass black holes that would, however, not merge within a Hubble time \textit{in an isolated binary}.

\begin{figure}
    \centering
    \includegraphics[width=0.5\textwidth]{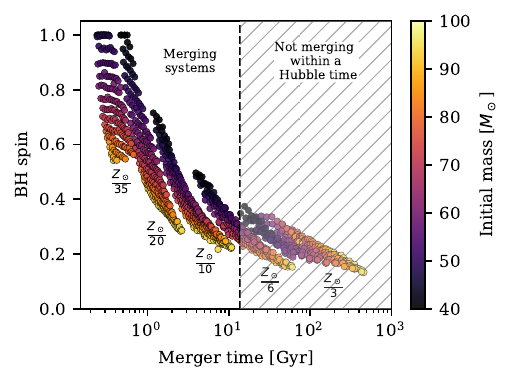}
    \caption{Final BH spin and merger times for isolated CHE models at different metallicities. At low metallicity systems reach high spins and all system merge. With increasing metallicity, the spins decrease, but the merger times increase until at a metallicity of $Z_\odot/3$ none of the systems would merge within a Hubble time. For clarity, we only show systems with masses below $100\, M_\odot$.}
    \label{fig:spin_mergertime}
\end{figure}

The effect of metallicity is shown in Fig.~\ref{fig:spin_mergertime}, which displays the spins and merger times of the BHs formed from the CHE models at different metallicities. This figure illustrates the 
following main trend:

\begin{itemize}
    \item At the lowest metallicities, the CHE models produce BH mergers with short delay times on the order of a few hundred Myr and exhibit high spins ($>0.5$).
    
    \item Models at the upper end of the metallicity range exhibit significantly lower spins ($<0.4$), but none of these systems merge within a Hubble time.
\end{itemize}

The lower spins and longer merger times are a consequence of enhanced stellar winds at higher metallicities, which leads to greater mass loss in the metal-rich models \citep{Vink_2005}. As a result: (i) the angular momentum loss reduces the stellar and BH spins, and (ii) the orbit widens, increasing the merger time. In this work we refer to the dimensionless spin magnitude of each BH as the ``BH spin'', for brevity.

Figure~\ref{fig:spin_mergertime} illustrates that the CHE channel, when occurring in isolation, is restricted to low-metallicity environments and predicts relatively high spins. In the analysis, we will investigate whether the presence of a tertiary companion might reduce the merger time for the highest metallicity considered in the grid. Since the majority of star formation in cosmic history occurs at higher metallicities, finding enhanced merger rates in this regime can significantly impact the predictions of the channel.

\subsection{Three-body dynamics}
The von Zeipel-Lidov-Kozai (ZLK) mechanism \citep{Zeipel_1910, Lidov_1962, Kozai_1962} is a secular effect that can play an important role in the dynamics of a triple system. In a hierarchical triple, where the third star is at a comparatively large distance from the other two stars, the system can be described as an inner binary (consisting of the two close stars with masses $m_1$ and $m_2$) and an outer binary (consisting of the third star of mass $m_3$ that orbits the center of mass of the inner binary with total mass $m_1 + m_2$). We closely follow the methodology presented in \cite{Vigna_2025}.

Due to gravitational torques exerted by the tertiary companion, the eccentricity of the inner binary, as well as the inclination between the inner and outer orbits vary periodically. These are known as ZLK oscillations and they have been extensively studied in the literature \citep[e.g.][]{Naoz_2016}. In the following, we summarize their main effects on the evolution of triple systems. 

In the quadrupole approximation, the ZLK oscillations act on a timescale given by \citep{Kiseleva_1998,Kinoshita_1999, Liu_2015}:
\begin{equation}
     t_\textnormal{ZLK} = \frac{(m_1+m_2)^{1/2}a_\textnormal{out}^3}{G^{1/2}\,m_3\,a_\textnormal{in}^{3/2}} (1 - e_\textnormal{out}^2)^{3/2} \, ,
    \label{eq:LK}
\end{equation}
where $a_\textnormal{in}$ and $a_\textnormal{out}$ are the semi-major axes of the inner and outer orbits, and $e_\textnormal{out}$ is the eccentricity of the outer orbit. Under the influence of the tertiary, the induced oscillations in the inner binary's eccentricity reduce their pericenter distance. In some cases, this can even lead to a merger of the inner binary stars. The maximum eccentricity $e_\textnormal{max}$ that can be achieved through the ZLK mechanism in the classical approximation is given by:

\begin{equation}
    e_\textnormal{max, classic} = \sqrt{1 - \frac{5}{3} \cos^2{i}} \, , 
    \label{eq:emax_classic}
\end{equation}
where $i$ is the initial mutual inclination of the inner and outer orbits. This equation shows that ZLK oscillations are the strongest at a mutual inclination of 90\textdegree, with the ZLK mechanism being active between 39.2\textdegree \,and 140\textdegree. 

The above equation is derived using the test-particle approximation, under the assumption that one of the inner stars is a massless test particle, as well as averaging the Hamiltonian over the inner and outer orbits, which is known as the double-averaging (DA) procedure. The interaction potential is also truncated at the leading quadrupole order, neglecting higher-order terms such as the octupole term. This is not an issue in our case, as the octupole term does not play a significant role for equal-mass inner binaries, and the quadrupole approximation remains valid \citep{Liu_2015}. 

However, the test-particle approximation and the DA assumption are not always valid and the eccentricity calculation requires corrections in the regimes where these assumptions break down. Moreover, there are short-range forces, such as tides or general relativistic (GR) effects, that can induce an additional source of apsidal precession in the inner binary. In some systems, these short-range forces might act on shorter timescales than the secular ZLK oscillations and they can interfere with its resonances, therefore suppressing or even entirely quenching the ZLK mechanism \citep{Fabrycky_2007, Liu_2015, Toonen_2016, Bataille_2018}. The rotationally driven oblateness of the stars can also induce precession, but we neglect this effect since it is subdominant compared to the tidal precession \citep{Liu_2015}. 

The tidal apsidal precession timescale is given by \citep{Smeyers_2001, Liu_2015}:
\begin{equation} 
t_\textnormal{tides} = \frac{1}{15}\frac{1}{k}\frac{m_1}{m_2}\frac{1}{\sqrt{G(m_1+m_2)}}\frac{a_\textnormal{in}^{13/2}}{r_1^5} \, , 
\label{eq:tides}
\end{equation}
where  $r_1$ is the radius of the primary star in the inner binary and $k$ ($= 0.0144$ for a main-sequence star) is the apsidal motion constant, which relates to the Love number $k_L$ through $k_L = 2 \cdot k$.
Tides can also circularize the orbit, which can quench the ZLK oscillations. However, since the apsidal precession acts on timescales much shorter than the circularization process, we only focus on the former mechanism.

The timescale of the apsidal precession induced by GR effects is given by \citep{Blaes_2002, Miller_2002}:  
\begin{equation}
    t_\textnormal{GR} = \frac{2\pi c^2 }{3[G(m_1+m_2)]^{3/2}} a_\textnormal{in}^{5/2} \, , 
    \label{eq:GR}
\end{equation}
where $c$ is the speed of light.  

In order to understand whether the ZLK mechanism has an influence on the inner system, we will have to compare the timescales on which it acts (Eq.~\ref{eq:LK}) with the timescales of the short-range forces (Eqs.~\ref{eq:tides},~\ref{eq:GR}). If the short-range forces act on a much shorter timescale than the ZLK mechanism, such that 
\begin{equation}
    \textnormal{min}(t_\textnormal{tides}, t_\textnormal{GR}) \ll t_\textnormal{ZLK} \, , 
\end{equation}
ZLK oscillations will be quenched. 

When the double-averaging assumption and the test-particle limit are relaxed and when corrections due to the short-forces are included, Eq.~\ref{eq:emax_classic} becomes more complicated. The maximum eccentricity is then approximated by the following equation, using the analytical expression from \cite{Mangipudi_2022}:
\begin{equation}
\begin{split}
    0 = -5 g_{\max}^2 + j_{\min}^2 \left( 3 - \cos i \eta - \frac{\eta^2}{4} e_{\max}^2 \right) \\
    \quad + \frac{3 \epsilon_{\text{SA}}}{8} \left[ -\frac{j_{\min}^2 \cos^3 i}{e_{\max}^2} - \frac{\eta j_{\min}^2}{2} - 9 j_{\min}^2 g_{\max} \right] \\
    \quad + g_{\max}^3 \left( \frac{1}{e_{\max}^2} - 16 \right) + \frac{8 \epsilon_{\text{GR}} j_{\min}}{3 e_{\max}^2} (j_{\min} - 1)\, ,
    \label{eq:emax_cor}  
\end{split}
\end{equation}
with $g_{\max} = \cos i + \eta e_{\max}^2/2$ and $j_\textnormal{min} = \sqrt{1-e_\textnormal{max}^2}$, where $\eta $ is the ratio of the circular inner angular momentum to the outer angular momentum and quantifies the level of hierarchy, $\epsilon_\text{SA}$ is the single-averaging term correcting for the DA approximation and $\epsilon_\text{GR}$ is the correcting term for GR effects.
There are additional correction factors that can be included but that are not relevant for the current project. For example, we do not include any correction for tidal forces from the pre-collapse stars, since we only explicitly compute the maximum eccentricity after BH formation.

In the case that the inner binary avoids a stellar merger and the components form BHs, the oscillations in eccentricity induced by the tertiary companion lead to closer passages of the BHs and, thus, increase their chance of merging. Under the influence of the ZLK mechanism, the merger time of the BBH system, defined as the time between BH formation and merger, is approximately reduced to \citep{Liu_2018}:

\begin{equation}
    \tau_\textnormal{merger} = \tau_\textnormal{isolation}(1 - e_\textnormal{max}^2)^3 \,\, , 
\end{equation}
where $t_\textnormal{isolation}$ is the merger time that the inner binary would need in isolation without the influence from a tertiary companion \citet{Peters_1963}. 

This shows how the eccentric behavior induced by the ZLK mechanism can shorten the merger time of the BHs, and potentially trigger a merger in a system that would not have merged otherwise in an isolated binary. 

Since we use analytical prescriptions to model the evolution of the triple systems, we restrict our analysis to dynamically stable and hierarchical configurations. If the tertiary companion is too close to the inner binary, the hierarchical configuration breaks down, leading to chaotic three-body interactions that may result in component exchanges, mergers or ejections \citep{Kiseleva_1994, Perets_2012, Toonen_2022}. Such dynamically unstable systems, as well as systems for which the secular approximation breaks down, require direct N-body integrations, which are beyond the scope of this work. For the threshold between dynamically stable and unstable systems, we use the criterion of \citep{Mardling_2001} given by:
\begin{equation}
\frac{a_{\mathrm{out}}}{a_{\mathrm{in}}}
= \frac{2.8}{1 - e_{\mathrm{out}}}
\left[\frac{(1 + q_{\mathrm{out}})(1 + e_{\mathrm{out}})}
{\sqrt{1 - e_{\mathrm{out}}}}
\right]^{2/5}\left(1-\frac{0.3i}{\pi} \right)
\label{eq:marding}
\end{equation}
where $a_\mathrm{out}$ and $a_\mathrm{in}$ are the outer and inner separations, $q_\mathrm{out} = m_3/(m_1+m_2)$, $e_\mathrm{out}$ is the eccentricity of the outer orbit and $i$ the mutual inclination between the inner and outer orbits (see also \citealt{Vynatheya_2022} for an updated prescription). 

In the following analysis, we use the grid of detailed stellar models for the inner CHE binary system, while the dynamical effects of the tertiary companion are evaluated using the analytical prescriptions described above.

\section{Results}
\label{sec:results}

To explore the role of a tertiary companion on the CHE channel we focus on the higher metallicity grid at \makebox{($ Z_\odot/3$)}, where none of the systems merge as isolated binaries. We start by analyzing a representative triple system in Sect.~\ref{sec:3.1}, and then show how the ZLK-induced reduction in merger times depends on the mutual inclination and outer separation of the triple systems in Sect.~\ref{sec:3.2}. In Sect. ~\ref{sec:3.3}, we explore how the ZLK-active window that leads to mergers in the inner BHs depends on the parameters of the system, such as the component masses, inner and outer separations. At last, we examine the impact of a fixed tertiary companion on the full CHE grid in Sect.~\ref{sec:3.4}. 

\subsection{Case study: A representative inner CHE binary in a triple system with varying outer separation}
\label{sec:3.1}

\begin{figure}
    \centering
    \hspace*{-0.5cm}
    \includegraphics[width=0.5 \textwidth]{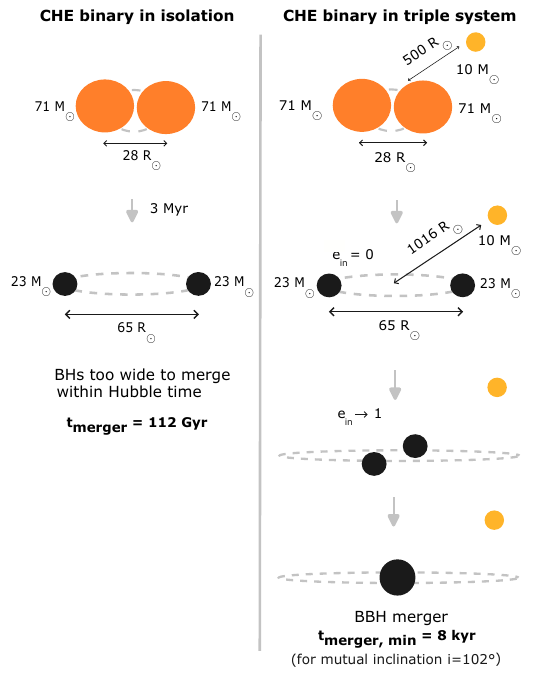}
    \caption{Schematic representation of the CHE binary channel in isolation (left panel), where the final low-spin BHs are too wide to merge within a Hubble time, and the same CHE binary in a triple system (right panel), where the eccentric oscillations induced by the tertiary companion drive the inner binary to a merger. }
    \label{fig:triple_system}
\end{figure}

\begin{table}[]
    \centering
\begin{tabular}{l|c|c|c}
    & & Initial & Final \\
     \hline
     \hline
Inner masses & $m_{1,2}$ [$M_\odot$] & 71 & 23 \\
Tertiary mass & $m_3$ [$M_\odot$] & 10 & 10 \\
Inner separation & $a_\mathrm{in}$ [$R_\odot$] & 28 & 65 \\
Outer separation & $a_\mathrm{out}$ [$R_\odot$] & 500 & 1016 \\
Inner eccentricity & $e_{\rm{in}}$ & 0 & 0.998 \\
Outer eccentricity & $e_{\rm{out}}$ & 0.67 &   \\
Mutual inclination & i [$^\circ$] & 102 & \\ 
\hline
Tidal timescale & $t_\mathrm{tides}$ [yr] & 0.1 & -- \\
GR timescale & $t_\mathrm{GR}$ [yr] & 118 & 5246 \\
ZLK timescale & $t_\mathrm{ZLK}$ [yr] & 21 & 28 \\
\hline
Merger time & $t_\mathrm{merger}$ & 112 Gyr & 8 kyr \\
ZLK mechanism&  &quenched& active\\
\end{tabular}
\vspace{0.3cm}
\hspace{-0.2cm}
\caption{Properties of the representative triple system for the initial configuration, where the inner system consists of two main-sequence CHE stars, and the final configuration, where the inner stars have formed a binary black hole system. The comparison of the timescales for the different processes that induce precession in the inner binary shows that the ZLK mechanism, which is quenched by tides during the main-sequence, becomes active once the stars form BHs. This induces a merger in the inner BHs within a Hubble time by reducing the GW inspiral time from the 112 Gyr they would need in isolation to 8 kyr, for the chosen configuration.}
\label{tab:system}
\end{table}

\begin{figure*}
    \centering
    \includegraphics[]{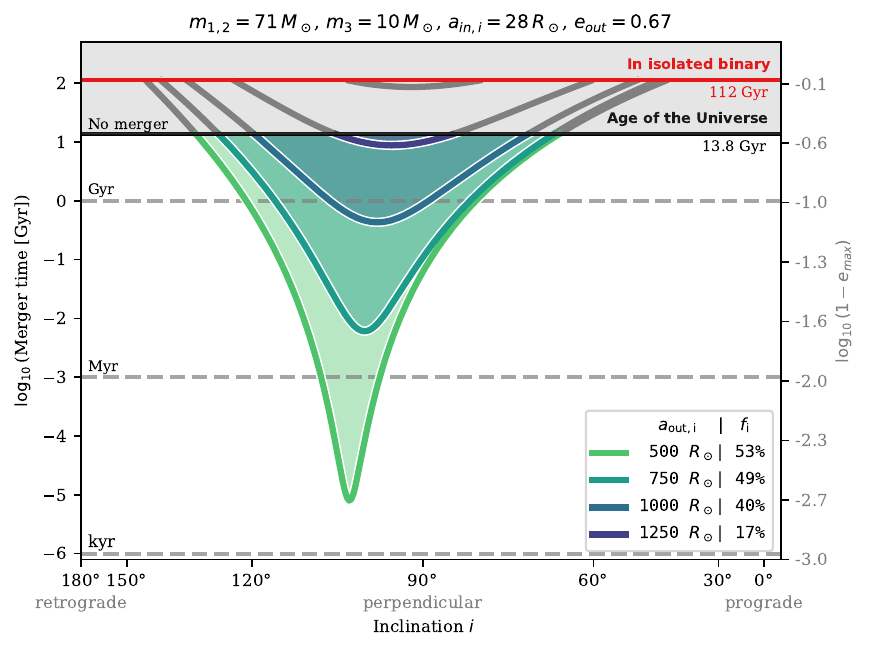}
   \caption{Merger times from GW emission of the representative inner BH binary under the influence of a $10\, M_\odot$ tertiary for different mutual inclinations, $\cos{(i)}$, and initial outer orbital separations, $a_\text{out,i}$. For moderately high inclinations, ZLK oscillations induce mergers within the age of the Universe in the inner BH binary system that would have otherwise required 112 Gyr in isolation. The strongest effect occurs for compact triples ($a_\text{out,i} = 500\, R_\odot$), where very prompt ($t_\textnormal{merger} < 1 \, \text{Myr}$) highly eccentric ($e_\mathrm{max} > 0.998$) mergers are possible. At larger outer separations, GR precession becomes increasingly effective at suppressing the ZLK oscillations, until the inner binary approaches isolated evolution once $a_\text{out,i} \gtrsim 1250\, R_\odot$. For each outer separation, the legend indicates the fraction of inclinations that merge within the age of the Universe, $f_\text{i}$, assuming random inclinations.} 
    \label{fig:merger_time_e_i}
\end{figure*}

 We first discuss a representative triple system, in which the inner CHE binary leads to the formation of two low-spin black holes that do not merge within a Hubble time in isolation. In Fig.~\ref{fig:triple_system} we illustrate the evolution of the binary system in isolation compared to its possible evolutionary path in the presence of a third object. The properties of the triple system are displayed in Table~\ref{tab:system}. 

The inner system consists of a CHE binary with component masses $m_1 = m_2 = 71 \, M_\odot$, and it was selected from the model grid with the highest metallicity \makebox{($ Z_\odot/3$)} described in Section~\ref{sec:che_models}. At this metallicity, stellar winds lead to significant mass loss, with about 2/3 of the system's initial mass lost by the time the BHs form, which causes the orbit to widen by more than a factor of 2. Because of the angular momentum loss through stellar winds, the final BHs have relatively low spins of 0.2. However, the system becomes too wide to merge within a Hubble time, with the BHs taking more than 100 Gyr to merge in an isolated binary.

In this example, we choose a tertiary companion of $10\, M_\odot$ at an initial separation of $a_{\mathrm{out,i}} = 500 \, R_\odot$. We choose an outer eccentricity of $e_{\rm{out}} = 0.67$, corresponding to the mean of a thermal distribution, $p(e_{\rm{out}})=2e_{\rm{out}}$, which is consistent with the observed outer-eccentricity distribution of hierarchical triples \makebox{\citep[e.g.][]{Heggie_1975}}. We neglect the evolution of the tertiary companion, since the more massive inner stars evolve on much shorter timescales (3 Myr, compared to the tertiary's lifetime of 21 Myr). 

In the initial configuration, the inner binary is in a very tight orbit with a separation of only $28\, R_\odot$, while each star has a radius of approximately $10\,R_\odot$. This close proximity leads to strong tidal interactions, which effectively shield the system from dynamical perturbations induced by the tertiary companion. In particular, the tidal apsidal precession acts on shorter timescales than the ZLK mechanism ($t_\mathrm{tides} = 0.1 \text{ yr} <  t_\mathrm{ZLK} = 21 \text{ yr}$). As a result, ZLK oscillations are quenched during the main-sequence phase of the inner binary.

\begin{figure*}
    \centering
    \includegraphics[width=\linewidth]{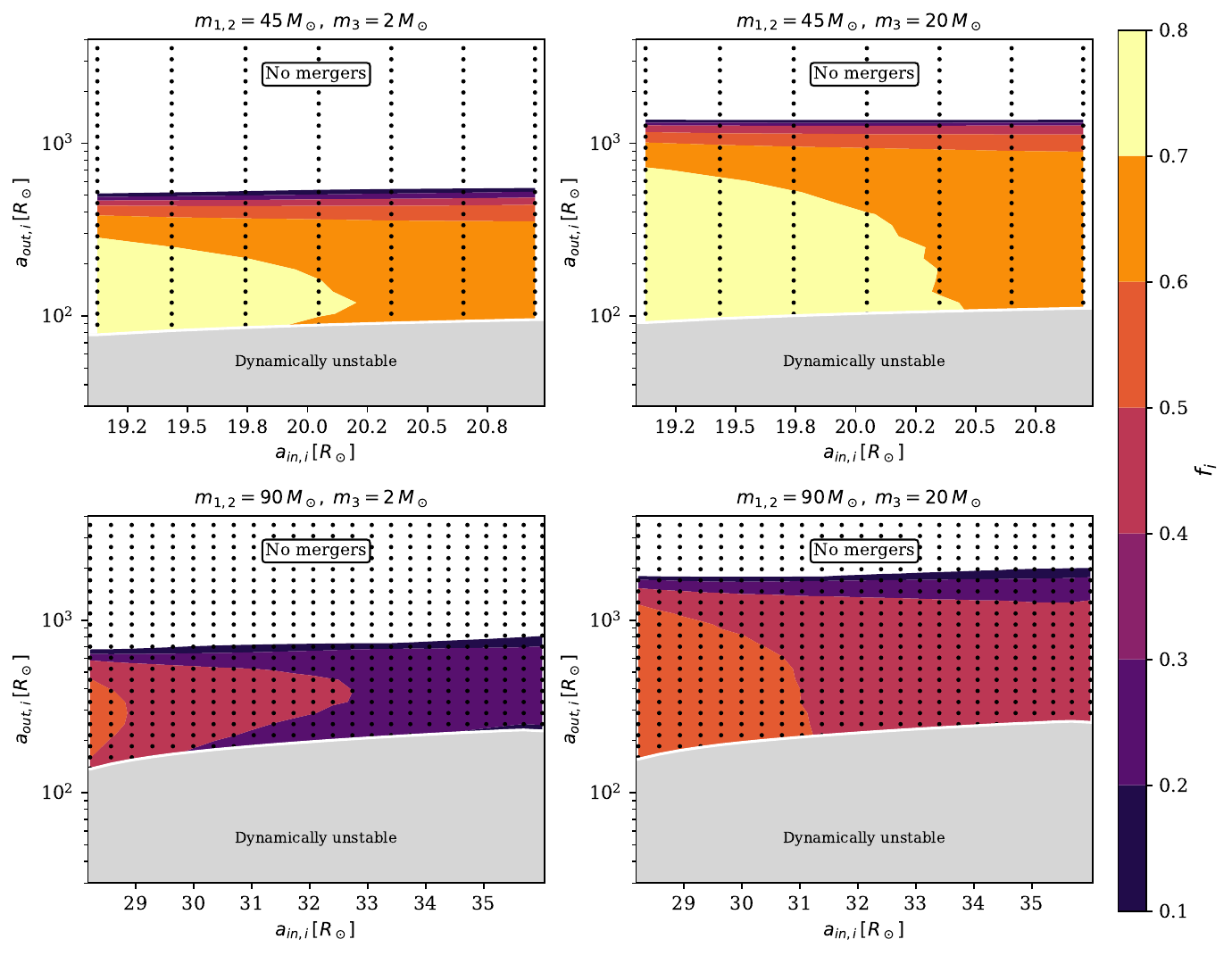}
    \caption{Fraction of systems that merge within a Hubble time assuming random orientations, $f_i$, as a function of the initial inner and outer separations for different triple configurations. The fractions are indicated by the color and is interpolated between the black grid points that indicate the modeled systems. The inner separations are chosen from the CHE window in the binary grid. The colors, interpolated between the black grid points, indicates the fraction of mutual inclinations that lead to a merger of the inner binary within a Hubble time through the ZLK mechanism. The panels show inner binaries with $m_{1,2}=45\,M_\odot$ (top row) and $90\,M_\odot$ (bottom row) in triple systems with tertiaries of $m_3=2\,M_\odot$ (left column) and $20\,M_\odot$ (right column). Grey regions denote dynamically unstable configurations and the white regions denote systems, in which GR precession suppresses ZLK oscillations and prevents mergers. }
    \label{fig:merger_fraction}
\end{figure*}

However, once the inner stars form BHs, tidal forces vanish and the GR effects alone are not strong enough to quench the ZLK oscillations \makebox{($t_\mathrm{GR} = 5\cdot 10^3 \text{ yr} > t_\mathrm{ZLK} = 28$ yr)}. Consequently, the ZLK mechanism starts to play an active role in the system's dynamics. This leads to a prompt binary black hole merger on a timescale of 8 kyr for a mutual inclination between the inner and outer orbits of $i=102^\circ$, which was chosen to maximize the ZLK effect for this particular system.

\subsection{Dependency on the inclination and outer separation}
\label{sec:3.2}
Whether the ZLK mechanism can efficiently drive a merger depends on the inclination between the inner and outer orbits and the outer separation of the tertiary. To explore this dependence, Fig.~\ref{fig:merger_time_e_i} shows the merger time of the representative low-spin binary black hole system in the presence of a $10\,M_\odot$ tertiary for different outer separations and mutual inclinations. 

The closer the tertiary, the stronger the eccentricity excitation and the shorter the merger time. In compact triples with an initial outer separation of $a_{\rm{out,i}} = 500 \, R_\odot$, the tertiary drives mergers in the inner BH binary within a Hubble time for 53\% of the mutual inclinations, spanning approximately $65^\circ \lesssim i \lesssim130^\circ$. The most extreme evolution occurs near $i \approx100^\circ$, where the ZLK oscillations drive the inner binary to a highly eccentric orbit ($e_{\rm{max}}\approx0.998$), resulting in a very prompt merger within $\sim10\,\mathrm{kyr}$. The location of this peak deviates from the classical prediction, where the ZLK mechanism is most efficient at an inclination of $i =90^\circ$, corresponding to perpendicular inner and outer orbits (see Eq.~\ref{eq:emax_classic}). This shift arises because the systems deviate from a perfectly hierarchical configuration, which is accounted for by the $\eta$ factor in Eq.~\ref{eq:emax_cor}, with this effect becoming more prominent the closer the tertiary gets. As the outer separation increases, the GR precession progressively suppresses the ZLK oscillations, narrowing the range of inclinations that produce mergers. For $a_{\rm{out,i}} = 1250 \, R_\odot$, only 17\% of inclinations produce merging BHs and they occur on relatively long timescales of $\sim10 \text{ Gyr}$.

In conclusion, a tertiary companion can efficiently drive ZLK-induced mergers in the representative inner CHE binary in systems with the right outer separation ($a_\textrm{out,i} = 500 - 1250 \, R_\odot$) and mutual inclination ($65^\circ \lesssim i \lesssim 130^\circ$). The dependence on the most relevant parameters will be further explored in the following section.

\subsection{Merger fraction across the parameter space}
\label{sec:3.3}

The ZLK mechanism depends on the inner and outer separation of the triple system, the component masses, mutual inclination and outer eccentricity. In Fig.~\ref{fig:merger_fraction} we explore the relevant parameter space across different triple configurations. From the metallicity grid at $Z_\odot/3$ shown in Fig.~\ref{fig:period_mass}, we choose inner systems with masses $m_1 = m_2 = 45\,M_\odot \text{ and } 90 \, M_\odot$ (rows in Fig.~\ref{fig:merger_fraction}) with different initial separations across the CHE window; note that none of these systems would merge as isolated binaries. We then assume triple configurations with a tertiary with masses taken to be $m_3=2\, M_\odot \text{ and } \,20\, M_\odot$ (columns in Fig.~\ref{fig:merger_fraction}), and vary the outer separation between $a_{\rm{out,i}} = 30-3\,000 R_\odot$. 

For each triple configuration in the grid, indicated by the black points, we sample the mutual inclinations isotropically, i.e, uniformly in $\cos{i}$, and solve Eq.~\ref{eq:emax_cor} to determine the maximum eccentricity that can we induced through the ZLK mechanism after BH formation. We then compute the resulting merger time and determine the fraction of angles that lead to mergers within a Hubble time, $f_i$, and extrapolate between the points (indicated by the color). 

In Fig.~\ref{fig:merger_fraction} we can distinguish three main regions: if the tertiary is too close ($a_{\rm{out,i}} \lesssim 70-200\, R_\odot$) the system is dynamically unstable (gray) and we cannot analytically determine its evolution reliably. If the tertiary is too far away ($a_{\rm{out,i}} \gtrsim 500-1500\, R_\odot$), GR effects quench the ZLK mechanism and there are no mergers (white). In between these regions there is the parameter space where the ZLK mechanism is active, with the color indicating the percentage of mutual inclinations that lead to a merger in the inner CHE binary system. 

For a fixed inner binary mass, increasing the tertiary mass from $2\, M_\odot$ to $20\, M_\odot$ (compare the left and right column in Fig~\ref{fig:merger_fraction}) expands the upper bound of outer separations that lead to mergers from around $500\, R_\odot$ to $1500\, R_\odot$. This is expected since the higher the tertiary mass, the stronger the gravitational effect and the tertiary companion can be further away and have a similar impact on the inner binary.

For a fixed tertiary mass and different inner masses (compare the top and bottom row in Fig~\ref{fig:merger_fraction}), the upper bound in outer separations that lead to mergers does not change significantly. However, the merger fraction is different. Systems with an initial mass of $m_{{1,2}} = 45 \, M_\odot$ show higher merger fractions (70-80\%) compared to the $m_{{1,2}}=90\, M_\odot$ systems (40-60\%). This effect arises due to the larger separations and longer merger times in the final BBHs from the more massive systems. Not only do the $90\, M_\odot$ CHE binaries start with a larger initial separation, but due to stronger stellar winds the systems also widen more during the stellar lifetime, leaving BHs that are further apart. As a consequence, the merger times of the $90\, M_\odot$ models in isolated binaries ($\sim 100 \text{ Gyr}$) are much longer than for the $45\, M_\odot$ models ($\sim 15-30 \text{ Gyr}$), whose merger times are just above the Hubble time threshold. This means that for more massive inner binaries, the ZLK effect has to be much stronger, reducing the fraction of configurations that produce mergers within a Hubble time. 

Going from the 45 to the 90 $M_\odot$ inner models also shifts the dynamical stability boundary to higher values. Since triple systems with more unequal ratios between the tertiary and the inner binary system are more dynamically unstable, the tertiary has to be further apart for the system to be dynamically stable. The location of the dynamical stability boundary can also shift depending on the outer eccentricity (see Appendix~\ref{sec:app_merger_fraction}), for Fig.~\ref{fig:merger_fraction} we assume $e_\text{out}=0$ .

We conclude that there is a significant window in the investigated parameter space where the ZLK mechanism is active and a triple induces a merger within the age of the Universe in systems where the classical binary CHE channel would not operate. 

\begin{figure}
    \centering
    \includegraphics[width=0.5\textwidth]{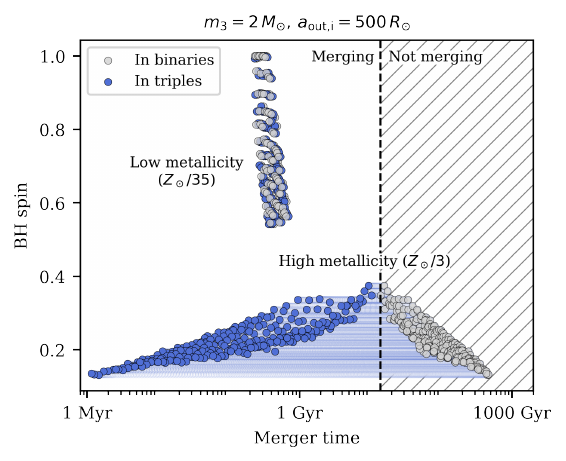}
    \setlength{\abovecaptionskip}{5pt}
    \caption{Spin and merger time of binary (gray) and triple (blue) systems at low ($Z_\odot/35$) and higher ($Z_\odot/3$) metallicity. We adopt a  mass of $m_3 = 2\, M_\odot$ and an initial outer separation of \makebox{$a_\textrm{out,i} = 500\, R_\odot$} for the tertiary. The transparent blue points show the merger times for different mutual inclinations, while the darker blue points indicate the minimum merger time achieved for the inclination that leads to the strongest ZLK effect. At the highest metallicity, the ZLK mechanism drives otherwise non-merging BBHs to coalescence, reducing the merger time from longer than a Hubble time to as short as $\sim 1 \,\textrm{Myr}$, depending on the mutual inclination. At the lowest metallicity, the final BHs are so close that GR effects quench the effect of the tertiary and the systems act as in isolation.
    } 
    \label{fig:spin_mergertime_triple}
\end{figure}

\subsection{Tertiary-induced mergers in the low-spin higher-metallicity CHE grid}
\label{sec:3.4}

To get an overview of the impact of a tertiary on the CHE grid, Fig.~\ref{fig:spin_mergertime_triple} compares the spins and merger times of the low ($Z_\odot/35$) and higher metallicity ($Z_\odot/3$) models in isolated binaries and triple systems (cf. Fig.~\ref{fig:spin_mergertime}). For the triple systems, we adopt a $2\, M_\odot$ outer companion at an initial separation of $500\,R_\odot$ and sample the mutual inclination between the inner and outer orbits isotopically. In binaries (gray), only the low-metallicity systems merge within a Hubble time. Triples (blue) change this picture by driving mergers in the higher-metallicity models, reducing the merger times from a hundred of Gyr to even 1 Myr, in the most extreme cases. This gives rise to a population of low-spin mergers with properties distinct from the high-spin isolated binaries, which will be analyzed more in detail in Section~\ref{sec:4}.

The spread in merger times in the higher-metallicity triple models is driven by the different mutual inclinations, which determine the strength of the ZLK mechanism. The systems with the lowest spins, which correspond to the most massive models (cf. Fig.~\ref{fig:spin_mergertime}), experience the most drastic reduction in merger time. These systems lose more mass through stellar winds, leading to the lower spins and also causing the inner orbit to widen more. This reduces the quenching effect of the GR precession on the ZLK oscillations, allowing stronger eccentricity excitations and faster mergers. This behavior is also apparent in Fig.~\ref{fig:merger_fraction} (left column). For the $m_{{1,2}} = 45 \, M_\odot$ models, systems with $a_\text{out,i} = 500 \,R_\odot$ lie very close to the boundary of the no-merger region, where GR precession dominates. Comparatively, the same outer separation lies well within the ZLK-active window for the $m_{{1,2}} = 90 \, M_\odot$ models, allowing the tertiary to efficiently drive mergers.  

The low-metallicity systems are essentially unaffected by the tertiary because they stay very compact and apsidal precession induced by GR effects completely quench the ZLK mechanism. This holds true for the chosen triple configuration. More massive or closer triple companions can, however, affect the merger times even in the low metallicity regime, as will be shown in the next section.

\section{Distribution of merging black holes}
\label{sec:4}

In the previous section, we show that the presence of a triple can enable mergers in low-spin systems that would otherwise not merge in isolated binaries. This raises the question of how the properties of merging BBHs in triples differ from those produced through isolated CHE binaries.  

A full exploration of this requires integrating over the cosmic star-formation history, which we plan to present in a follow-up study. Here, we explore the first consequences for the populations, based on toy models at fixed metallicities. To sample the binary and triple populations, we follow the approach of \citet{Moe_2017} and \citet{Toonen_2022}, with the modifications adopted here described in Appendix~\ref{sec:app_pop_synth}. 

Figure \ref{fig:distributions} shows the mass, spin and merger-time distributions of the merging BBHs from binaries and triples at two representatives metallicities ($Z_\odot/35$ and $Z_\odot/3$). For readability, we will refer to this as low and high metallicity, respectively. Even though the latter is not generally thought of as high metallicity, it is the highest metallicity available in the grid we use. Our proposed CHE channel in triples operates at these higher metallicities (shown in blue) and produces BHs with different population properties compared to the classical binary CHE channel (shown in pink), which only operates at lower metallicity. By opening up the channel to higher metallicities, the masses (top panel), as well as the spins (middle panel) of the merging BHs in triples are shifted to lower values. This is a direct consequence of enhanced stellar winds at higher metallicity, which remove mass and angular momentum from the progenitors stars. Moreover, triple dynamics significantly accelerates the merger times of the BBHs compared to binary systems that evolve in isolation (bottom panel). 

More quantitatively, for the metallicities chosen here as examples, we find that the masses (top panel) of high-metallicity triples (blue) peaks near $15\, M_\odot$, while low-metallicity binaries (pink) peak around $30\, M_\odot$. We expect that triples at even higher metallicities would lead to distributions that peak at even lower masses. This suggests that CHE in triples could have a promising contribution to the global peak at $10\,M_\odot$ from the GW data \citep{Ligo_2025}.

The spins experience an even more drastic shift (middle panel). High-metallicity triples (blue) exclusively lead to low-spin BHs below 0.4, with the distribution peaking around $0.25$. This contrast with the classical low-metallicity binary population (pink), which produces rapidly spinning BHs, with $45\%$ of the systems having extreme spins above $0.9$. The pile-up at 1 arises from highly spinning progenitors whose angular momentum exceeds that of a maximally spinning BH. In these cases, the resulting BH spin is capped at the Kerr limit of $1$.

For the binaries, there is also a small contribution at low spins, that accounts for $3\%$ of the population. These originate from progenitors above the pair-instability mass gap ($\gtrsim120\, M_\odot$). The significant spin-down in these systems is caused by extreme mass loss that these very massive progenitors experience. The lower spins in triples are more consistent with the overall BBHs population from the GW data, which is dominated by low-spin mergers \citep{Ligo_2025}. 

\begin{figure}
    \centering
\includegraphics[width=0.5\textwidth]{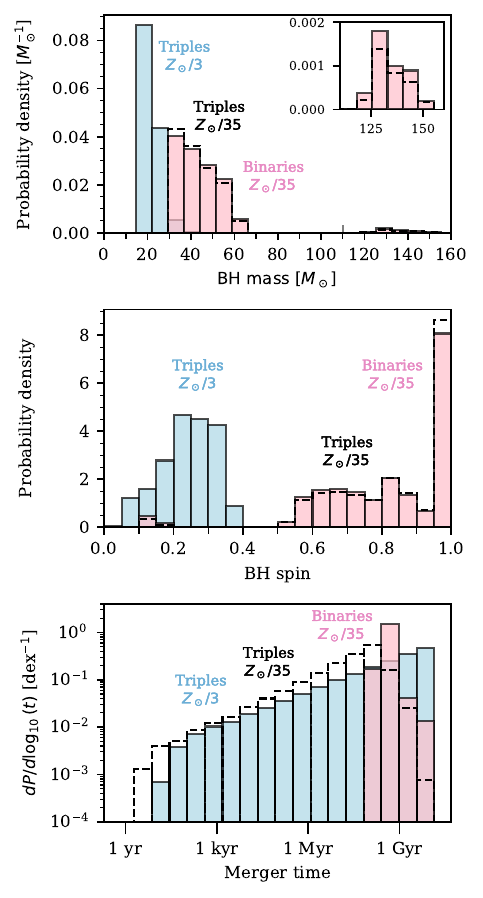}
\caption{Distributions of masses, spins and merger times of merging black holes. Comparison between CHE in triples at high metallicity ($Z_\odot/3$, blue), triples at low metallicity ($Z_\odot/35$, dashed black) and isolated CHE binaries at low metallicity ($Z_\odot/35$, pink).}
\label{fig:distributions}
\end{figure}

Triple dynamics shortens the merger times (bottom panel), giving rise to a broad range of timescales from Gyr to hundreds of years \citep[cf.][]{Antonini_2017, Toonen_2022, Dorozsmai_2023, Vigna_2025}. In contrast, binaries from the classical CHE scenario have a narrow distribution of merger times that peak just under 1 Gyr. Half of the high-metallicity triple population (blue) merges within 1 Gyr. These short delay times suggest that the triple merger rate traces the cosmic star-formation history sufficiently closely to contribute significantly to the local BBH merger rate. Moreover, 10\% of the population is found to merge within 1 Myr (and 1\% within 1 kyr). Events with such short merger timescales are of particular interest as potential sources of electromagnetic counterparts associated with GW events \citep[e.g.][]{Kelly_2017, Mink_2017, Stone_2017, Fragione_2019, Naoz_2025, Kummer_2025}, since the merger may occur while the surrounding environment still contains material expelled during the evolution of the progenitor stars.

It is instructive to also compare the two channels at the same metallicity. We can only do this for the low metallicity, where both channels operate. Comparing the binary (pink) and triple (black, dashed) population, shows that they share very similar mass and spin distributions. The tertiary thus does not change the intrinsic properties of the inner binary at a fixed metallicity. However, it does reduce the merger times drastically, as can be seen in the bottom panel, similarly as in the case of the high-metallicity models. Around 65\% of the low-metallicity triples merge within 100 Myr, while none of the binaries at the same metallicity do. Binaries only produce BBHs with longer merger times above 200 Myr.

Even though we do not track the evolution of the mutual inclinations and eccentricities of the systems, the triple-induced mergers are expected to show spin-orbit misalignment based on previous studies \citep{Liu_2018, Liu_2019, Antonini_2018, Su_2021, stegmann2026massspinpropertiesblackhole}. This differs from the expectation of orbitally-aligned spins from the classical binary CHE channel. Moreover, even though GW emission circularizes the orbit, triples can excite the inner system to such high eccentricities that a small fraction are expected to retain a non-zero eccentricity at merger that is measurable in the LIGO band \citep{Wen_2003, Antonini_2018, Liu_2019, Toonen_2022}.

In summary, we have shown that triples  can produce merging BHs with (i) lower masses, (ii) lower spins and (iii) a wide range of delay times including very prompt events of possible interest for EM counterparts. Furthermore, because triples open up the channel at high metallicity, we further expect the rates to increase. Although modeling, of the full population requires further work, it is promising that these properties are more in line with the majority of the observed BBH systems.

\section{Discussion}
\label{sec: discussion}

We conclude by highlighting several assumptions and caveats of our work. 

Recently, major efforts have led to an increasing understanding of the proprieties of Galactic triple systems \citep[e.g.][]{Sana_2014, Moe_2017, Bordier_2026}. These findings highlight the ubiquity of massive triples and the importance to consider them in GW formational pathways. Despite this progress, the detailed population properties of triples, in particular for very massive systems, remain poorly constrained. Moreover, it is unclear to what extent these properties apply to different metallicities.

Our analysis inherits the uncertainties from the underlying CHE binary models. The efficiency of rotational mixing in stellar interiors and the onset of CHE remains a topic of debate. On the other hand, these models might underestimate the internal mixing as they do not account for deformation through tides that could potentially enhance this effect \citep{Hastings_2020}. 

The treatment of stellar-wind mass loss is also a source of uncertainty. Even though there are updated wind prescriptions \citep[e.g.][]{Sabhahit_2023, Bjorklund_2021, Bjorklund_2023, Higgins_2023}, they were not available at the time the grids were computed. The mass and angular momentum lost through winds directly affects the properties of the resulting BHs, such as their final masses and spins, as well as the orbital separation that dictates which systems lead to mergers \citep{Brookshaw_1998, Schroder_2021}. 

The final masses and spins of the BHs also depend on the assumption we make about the BH formation. We assume direct collapse of the stellar progenitor with no natal kick. This assumption is supported by theoretical models which find that BH formation with very low kicks is natural in the BH mass range relevant for this paper \citep[e.g.][]{Janka+2024,Popov+2025}.
Observations of some BH binaries, such as Cygnus \mbox{X-1}, VFTS 243, and V404 Cyg, are also consistent with a direct-collapse BH formation scenario with no, or very low, natal kicks \cite[e.g.][]{Mirabel+2003, Shenar_2022, Vigna_2024, Burdge_2024, Willcox_2025}. However, the final stages of BH formation are complex and, in case of rapidly spinning progenitors, the outer layers may have enough rotational support to form an accretion disk instead of directly collapsing into a BH, leaving a less massive, slowly spinning BH behind \citep{Beloborodov_1999, Fuller_2022, Gottlieb_2025}.

We use analytical prescriptions to describe the triple dynamics instead of direct three-body simulations. This allows us to explore a larger population efficiently, but we cannot track the real time evolution of the systems. We also have to discard dynamically unstable systems, which could potentially be interesting. Their chaotic three-body interactions may produce mergers that we are missing in our analysis \citep{Perets_2012, Toonen_2022}. 

To compute the merger times of the binary black holes due to GW emission, we use the orbit-averaged prescription from \citet{Liu_2018}. Recent work by \citet{Fumagalli_2026} investigated the accuracy of the orbit averaging by comparing the merger-time estimates with non-orbit-averaged calculations. They find that the prescription of \citet{Liu_2018} remains valid for most systems, but can potentially overestimate the merger times for compact, highly-eccentric systems.  

For simplicity, we do not model the evolution of the tertiary companion. We do not account for mass loss of the tertiary, supernova explosions, natal kicks and the possibility of perturbing or disrupting the triple system. We also do not account for any interaction between the tertiary and the inner binary, such as mass transfer or common-envelope evolution during the tertiary's post-main-sequence evolution. These assumptions are justified as long as a low-mass tertiary companion evolves on a timescale much longer than the inner system and the time it takes for the BHs to merge. For higher-mass tertiaries, the analysis is complicated by eventual expansion of the tertiary companion. Mass-transfer episodes or common-envelope evolution could potentially harden the inner binary and even shorten the merger time of the BHs, but they could also lead to stellar mergers \citep{Hamers_2022, Kummer_2025}. 

Despite these limitations, our results demonstrate that the triple CHE channel is a viable pathway for binary black hole mergers, particularly opening up the channel to higher metallicities. This changes the predictions of the CHE channel, leading to lower-mass low-spin BHs than expected from the classical scenario. For a calculation of the BH merger rate of the triple CHE channel an integration over metallicity across cosmic star formation is necessary and is left for future work.

\section{Conclusion}
\label{sec:conclusion}

We have investigated how the presence of a tertiary companion affects the evolution and fate of a chemically homogeneously evolving inner binary.
Originally, this channel was proposed for isolated binaries, where it predicts distinct outcomes for merging black holes: high masses (around $35\,M_{\odot}$ and above) and large spins aligned with the orbital angular momentum. Although some observed events potentially match these properties, they are not representative of the majority of the binary black hole population. Moreover, these predictions are based on studies that ignore that most massive stars reside in triples. We explore triple CHE evolution, extending upon the studies presented in \citet{Dorozsmai_2023} and \citet{Vigna_2025}, by combining a large grid of detailed stellar evolution MESA models with an analytic treatment of three-body dynamics. 
We show that triple interactions can significantly change the predictions of this channel. Our main findings are:

\begin{enumerate}
    \item \textbf{Tertiaries open up the CHE channel at higher metallicities. } 
    The classical binary CHE channel works exclusively at low metallicity. At higher metallicity, stellar winds widen the binary too much for it to merge. We show, in agreement with earlier work, that a tertiary companion can excite high eccentricities in the inner binary of these systems through the ZLK mechanism, enabling them to merge within the age of the Universe. Tertiaries thus extend the parameter space for this channel to higher metallicity, where most of the cosmic star formation has occurred  \citep{Chruslinska_2024}, and thus increase the channel's contribution to the BBH merger population.

    \item \textbf{The triple CHE channel produces lower-spin black holes.}
    The classical binary CHE channel only operates at low metallicity and predicts highly spinning BHs. In contrast, the triple channel also operates at high metallicity, where stellar winds spin down the progenitors of the BHs. We find that the BH spins peak around $0.25$ in simulations at $Z_\odot/3$, the highest metallicity considered in this work. We expect spins to decrease further at even higher metallicity. This is promising since the observed BBH population is dominated by systems with low spins. 

    \item \textbf{The triple CHE channel produces lower-mass black holes.}
    The enhanced mass loss at higher metallicity also shifts the BH mass distribution towards lower masses. At the highest metallicity considered in this work, $Z_\odot/3$, the peak occurs around $15\,M_\odot$. At even higher metallicities, we expect the BH mass distribution to peak at even lower masses, potentially contributing to the global peak in the GW data at $\sim 10\,M_\odot$.

    \item \textbf{A tertiary companion drastically shortens delay times, enabling very prompt mergers.}
    As also found in earlier work, triples can drastically shorten the delay times. We find a significant population of prompt mergers, with half occurring within 1 Gyr. This implies that, at low redshifts, the events closely follow cosmic star formation. About 10\% of the events even merge within 1 Myr (1\% within 1 kyr) and are of interest as potential sources of electromagnetic counterparts.
 \end{enumerate}

Since observations show that the majority of stars in the mass range relevant to BBH mergers are found in triple systems or higher order multiples, studying massive binaries in isolation provides an incomplete picture. Our findings show that considering the effect of a triple companion on an inner CHE binary changes the predictions of the classical channel by enabling BBH mergers at higher metallicities. The triple CHE channel leads to (i) lower masses, (ii) lower spins and (iii) a wide range of delay times, including prompt events, compared to the classical CHE channel.  We further expect triple dynamics to lead to spin-orbit misalignment and potential residual eccentricities at merger.

Although our simulations are sensitive to uncertainties in the model assumptions and a full assessment will require integrating over cosmic time, our results clearly highlight the importance of accounting for triple dynamics when predicting the properties and delay times of BBH mergers. Further exploration of the role of triple dynamics and how it interplays with stellar and binary evolution will be crucial for understanding the formation and properties of merging BBHs. This is still a relatively unexplored yet crucial avenue, as the rapidly growing GW catalog, as well as next-generation detectors, provide increasingly precise constraints on the BBH population.

\begin{acknowledgements}
We thank Silvia Toonen, Pablo Marchant and Abinaya Rajamuthukumar for insightful discussions during various stages of the project. This research was supported in part by the National Science Foundation under NSF grant No. 2009131, and NSF PHY-2309135 to the Kavli Institute for Theoretical Physics (KITP). We acknowledge the DFG cluster of excellence ORIGINS funded by the Deutsche Forschungsgemeinschaft (DFG) under Germany’s Excellence Strategy-EXC-2094-390783311. A.V.-G. is supported by the Carlsberg Foundation, grant CF24-1046. This work made use of the following software packages: \texttt{astropy} \citep{astropy:2013,astropy:2018,astropy:2022,astropy_21262391}, \texttt{matplotlib } \citep{Hunter:2007}, \texttt{numpy} \citep{numpy}, \texttt{pandas} \citep{mckinney-proc-scipy-2010,pandas_21500199}, \texttt{python} \citep{python}, \texttt{scipy} \citep{2020SciPy-NMeth,scipy_22051554}, and \texttt{sympy} \citep{sympy_15292153}.  Software citation information aggregated using \texttt{\href{https://www.tomwagg.com/software-citation-station/}{The Software Citation Station}} \citep{software-citation-station-paper,software-citation-station-zenodo}.

\end{acknowledgements}

\bibliography{ref}
\bibliographystyle{aa}


    
\appendix
\nolinenumbers

\label{sec:appendix}

\section{Merger fraction}
\label{sec:app_merger_fraction}

\begin{figure*}
    \centering
\includegraphics[width=\linewidth]{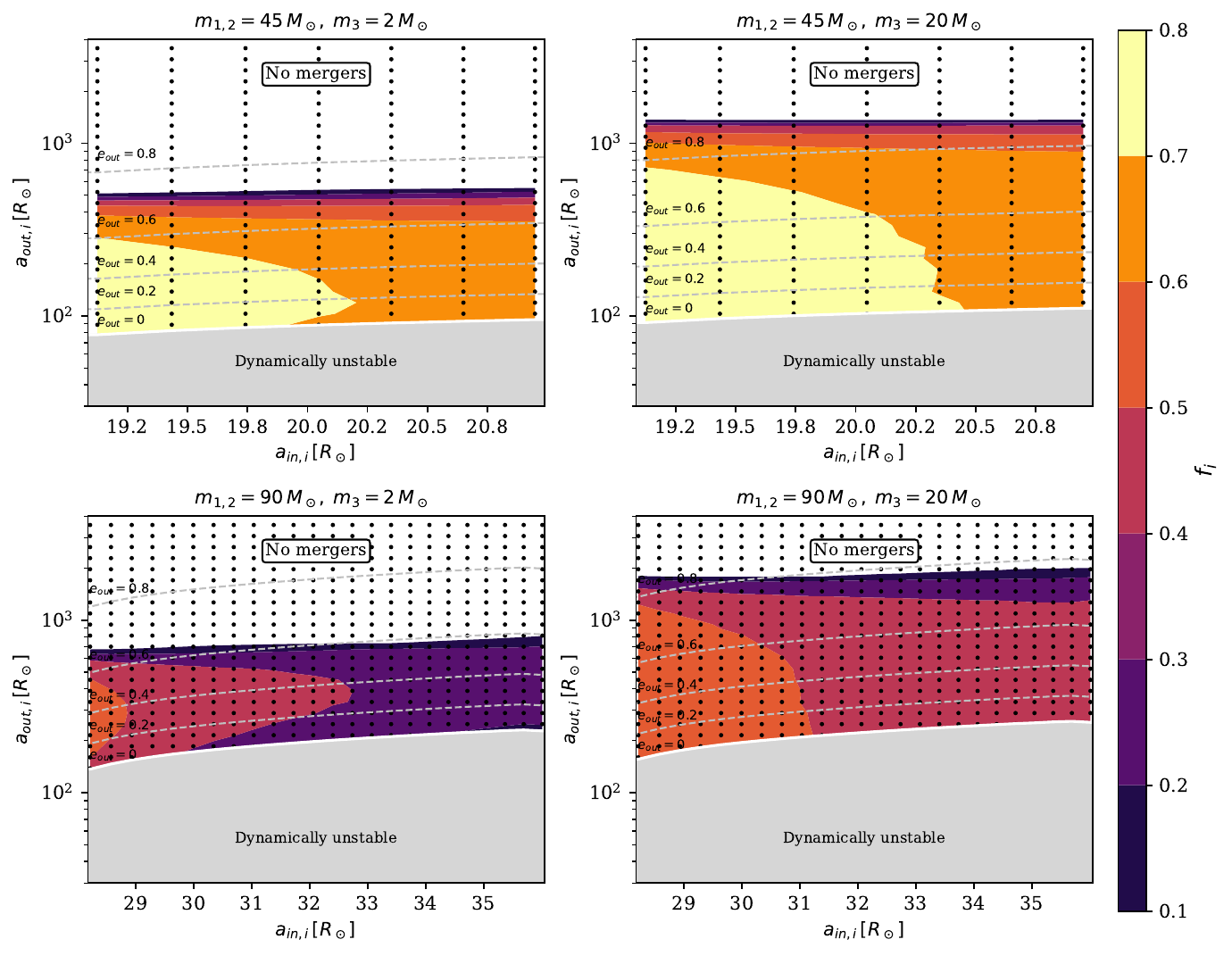}
    \caption{Same as Fig.~\ref{fig:merger_fraction}, but showing how the boundary of dynamical stability shifts for different outer eccentricities, $e_{\rm{out}}$. Higher outer eccentricities lead to less stable triple systems, requiring larger outer separations for the systems to be dynamically stable.}
    \label{fig:merger_fraction_eout}
\end{figure*}

In Fig.~\ref{fig:merger_fraction}, we show the fraction of systems that merge within a Hubble time for an isotropic distribution of mutual inclinations. We assume a circular outer orbit, $e_\text{out} = 0$. In reality, the outer eccentricities of triples are more likely to follow a thermal distribution \citep[e.g.][]{Heggie_1975}, with a mean outer eccentricity around $e_\text{out} = 0.67$. Changing the eccentricity does not impact the merger fractions, but it does shift the boundary of dynamical stability up. A higher outer eccentricity requires a larger outer separation for the system to remain stable, thereby reducing the range of stable configurations that lead to mergers. We show this in Fig. \ref{fig:merger_fraction_eout}.

\section{Population synthesis}
\label{sec:app_pop_synth}

We construct a toy population of hierarchical triple systems via Monte Carlo sampling, following an approach similar to \citet{Toonen_2022} with modifications to the parameter ranges. We sample the initial parameters from the following distributions:

\begin{itemize}

    \item \textbf{Inner-binary masses} The primary mass $m_1$ is drawn from a power-law distribution with a slope of $\alpha = -2.3$ in the range $20$--$300\,M_\odot$ \citep{Kroupa_2001}. We consider equal-mass binaries, $m_1= m_2$, following the assumptions in the detailed CHE MESA models that we base our population on.
    
    \item \textbf{Tertiary mass} We sample the outer mass ratio from a flat distribution, \begin{equation}
    q_\text{out} \sim \mathcal{U}(1/(m_1+m2), 8/(m_1+m_2)),
    \end{equation}
    with $q_\text{out}=m_3/(m_1+m_2)$, such that the tertiary mass $m_3$ is restricted to a range between  $1 - 8\,M_\odot$. We neglect higher mass companions to avoid the complexities associated with tertiary evolution and potential system disruption due to supernova kicks that we do not consider here.

    \item \textbf{Orbital separations} The inner and outer semi-major axes are drawn independently from log-uniform distributions,
    \begin{equation}
        a_{\rm in} \in [15, 100]\,R_\odot, \quad
        a_{\rm out} \in [20, 10^4]\,R_\odot.
    \end{equation}
    
        \item \textbf{Mutual inclination:} The inclination is sampled isotropically by drawing
    \begin{equation}
        \cos i \sim \mathcal{U}(-1, 1).
    \end{equation}
    
    \item \textbf{Eccentricities} 
    The outer eccentricity is drawn from a thermal distribution \citep[e.g.][]{Heggie_1975},
    \begin{equation}
    p(e_{\rm out}) = 2e_{\rm out},
    \end{equation}
over the range $ e_{\rm out}  \in [0, 1]$. The inner orbit is initialized as circular, $e_{\rm in} = 0$, as strong tidal forces between the stellar progenitors circularize the orbit.
\end{itemize}

We sample triple systems as described above and discard systems that are dynamically unstable according to the criterion described in Eq.~\ref{eq:marding}. Dynamically unstable systems are potentially interesting for prompt mergers. However, to follow their evolution requires N-body integrations, which are beyond the scope of the present study. For dynamically stable systems, we determine whether the inner binary undergoes CHE by matching the sampled parameters $(m_1, a_{\rm in})$ to the precomputed CHE grid. When a match is found, we extract the corresponding final binary properties, including the final mass, the final inner separation, the isolated merger time, and the black-hole spin.
We then update the outer orbital separation to account for the mass loss of the inner system and re-evaluate the dynamical stability of the post-CHE configuration. We assess whether the tertiary might induce a stellar merger in the inner binary. If the ZLK mechanism acts on shorter timescales than the apsidal precession induced by tides during the main-sequence of the inner stars, we assume that the tertiary induces a stellar merger in the inner system and we discard it. 

Systems that remain stable and do not merge as stars are further evolved by computing the maximum eccentricity induced by ZLK oscillations. This maximum eccentricity is then used to estimate the corresponding merger time under ZLK-driven evolution. For systems that merge withing a Hubble time, we record the relevant properties, including the initial parameters, orbital configuration, final spin, and merger time. 
Since the ZLK mechanism is a secular process, it requires the tertiary to undergo multiple orbits for the eccentricity excitation to take place. To avoid cases of very prompt mergers, we require the merger time to satisfy the following condition, $t_\text{merger} > 100 \, p_\text{out}$, to ensure that the system experiences sufficient number of outer orbits ($p_\text{out}$) for the secular ZLK evolution to operate. 

We track the number of dynamically unstable systems, systems undergoing CHE, and systems merging within a Hubble time. We sample triples at a metallicity of $Z_\odot/3$ and $Z_\odot/35$. For the separation distributions adopted, at $Z_\odot/3$, out of $10^7$ sampled triple systems, around \makebox{50\%} are dynamically stable. Among the stable systems, \makebox{5\%} undergo CHE with one in seven CHE systems leading to a BBH merger, resulting in a total of  $\sim 35\,000$ simulated merging BBHs systems. At a metallicity of $Z_\odot/35$, out of $2\cdot10^7$ sampled triple systems, \makebox{50\%} are dynamically stable, of which \makebox{5\%} undergo CHE with one in ten CHE systems leading to a BBH merger, resulting in $\sim 50\,000$ merging BBHs systems.

The isolated binary population is sampled using the same distributions as for the inner binaries of the triple systems at a metallicity of $Z_\odot/3$. We keep the binaries that undergo CHE and record their masses, final spins and merger times. Out of $10^6$ sampled binary systems around $40\,000$ undergo CHE and all of them merge within a Hubble time.

\end{document}